\documentclass[twocolumn,showpacs,preprintnumbers,amsmath,amssymb,prb,superscriptaddress]{revtex4-1}
\usepackage{bbm}
\usepackage{mathrsfs}
\usepackage{graphicx}
\usepackage{dcolumn}
\usepackage{bm}
\usepackage{amsmath}
\usepackage{amsfonts}
\usepackage{color}
\usepackage{floatrow}

\begin{document}

\title{Antiferromagnetic Dirac semimetals in two dimensions}
\author{Jing Wang}
\affiliation{State Key Laboratory of Surface Physics and Department of Physics, Fudan University, Shanghai 200433, China}
\affiliation{Collaborative Innovation Center of Advanced Microstructures, Nanjing 210093, China}

\begin{abstract}
The search for symmetry-protected 2D Dirac semimetals analogous to graphene is important both for fundamental and practical interest. The 2D Dirac cones are protected by crystalline symmetries and magnetic ordering may destroy their robustness. Here we propose a general framework to classify stable 2D Dirac semimetals in spin-orbit coupled systems having the combined time-reversal and inversion symmetries, and show the existence of the stable Dirac points in 2D antiferromagnetic semimetals. Compared to 3D Dirac semimetals which fall into two distinct classes, Dirac semimetals in 2D with combined time-reversal and inversion symmetries belongs to single class which is closely related to the nonsymmorphic space group symmetries. We further provide a concrete model in antiferromagnetic semimetals which supports symmetry-protected 2D Dirac points. The symmetry breaking in such systems leads to 2D chiral topological states such as quantum anomalous Hall insulator and chiral topological superconductor phases.
\end{abstract}

\date{\today}

\pacs{
        73.22.-f  % Electronic structure of nanoscale materials and related systems
        02.20.-a  % Group theory
        73.43.-f  % Quantum Hall effects
      }

\maketitle

\section{Introduction}
\label{introduction}

The discovery of the time-reversal invariant topological insulators~\cite{hasan2010,qi2011} greatly inspired the study of symmetry-protected Dirac semimetals (DSMs)~\cite{young2012,wangzj2012,wangzj2013,young2014,yang2014,liu2014a,liu2014b,borisenko2014,xiong2015,wieder2016,bradlyn2016,watanabe2016}. In a DSM, the low-energy physics is well described by pseudorelativistic Dirac fermions with linear energy dispersions along all momentum directions around Dirac points (DPs) in the Brillouin zone (BZ). To guarantee the robustness of three-dimensional (3D) DPs, the crystalline symmetry protection is a necessitate~\cite{yang2014}, similar to the 2D DPs in graphene~\cite{geim2009}. The 3D DSM materials~\cite{young2012,young2014,wangzj2012,wangzj2013,yang2014,liu2014a,liu2014b,borisenko2014,xiong2015} have both the time-reversal symmetry $\mathcal{T}$ and inversion symmetry $\mathcal{P}$. The $\mathcal{T}$ or $\mathcal{P}$ breaking in general leads to Weyl semimetals~\cite{murakami2007,wan2011,xu2011,burkov2011,liu2014,huang2015,weng2015,xu2015,lv2015,bernevig2015,sun2015,ruan2016}, where the DPs split into Weyl points. However, under certain conditions, such 3D DPs still remain stable with $\mathcal{T}$-broken~\cite{tang2016}. This motivates us to ask whether stable DPs could exist in 2D in the absence of $\mathcal{T}$ and $\mathcal{P}$. The positive answer to this question will lead to many interesting physical phenomena and novel topological states, as expected from the great success in the field of graphene.

Recently, the concept of 2D DSMs in the presence of spin-orbit coupling (SOC) has been introduced by Young and Kane~\cite{young2015}, where the nonsymmorphic space group symmetries play a key role~\cite{young2015,wieder2016b,chen2014,venderbos2016,zhao2016}. In this paper we study the 2D Dirac fermions in antiferromagnetic (AFM) semimetals. A general framework is proposed to classify stable 2D DSMs in SOC systems with $\mathcal{PT}$ symmetry. In sharp contrast to $\mathcal{T}$-invariant 3D DSMs which fall into two distinct classes~\cite{yang2014}, DSMs in 2D with $\mathcal{PT}$ symmetry only have one class. The DPs of 2D DSMs reside at the BZ boundary which are protected by the nonsymmorphic space group symmetries. We further provide a tight-binding model in AFM semimetals which supports symmetry-protected 2D DPs and symmetry breaking in such systems leads to exotic chiral topological states in 2D. We conclude with a brief discussion on the possible material venues for such phases.

The 2D DP with fourfold degeneracy is generated when two doubly-degenerate energy bands avoidably cross. The most common symmetry for crystalline solids to have double degeneracy of bands is $\mathcal{PT}$ symmetry satisfying $(\mathcal{PT})^2=-1$, which includes two cases with both $\mathcal{T}$, $\mathcal{P}$ conserved and $\mathcal{T}$, $\mathcal{P}$ broken. Under this condition, the low-energy physics can be described by the minimal four-band effective Hamiltonian,
\begin{equation}\label{minimal}
\mathcal{H}(\mathbf{k}) = \epsilon_0(\mathbf{k})+\sum\limits_{a=1}^5d_a(\mathbf{k})\Gamma_a,
\end{equation}
where $\Gamma_a$ are $4\times4$ Dirac $\Gamma$ matrices satisfying $\{\Gamma_a,\Gamma_b\}=2\delta_{ab}$, and the specific representation of $\Gamma_a$ depends on the crystalline symmetry. The particle-hole asymmetry term $\epsilon_0(\mathbf{k})$ is neglected for simplicity, $d_a(\mathbf{k})$ ($a=1,\ldots,5$) are real functions of $\mathbf{k}$, and $\mathbf{k}=(k_x,k_y)$. The energy spectrum is
\begin{equation}
E_\pm(\mathbf{k})=\pm\sqrt{\sum_{a=1}^5d_a^2(\mathbf{k})}.
\end{equation}
The DPs can be generated only when all $d_a(\mathbf{k})=0$ for certain $\mathbf{k}$ in the BZ. A simple case for this condition exists at the quantum critical point of the phase transition between a quantum spin Hall (QSH) and a normal insulator (NI) in the presence of $\mathcal{T}$ and $\mathcal{P}$~\cite{murakami2007,buttner2011}. Such DPs are accidental degeneracies and not robust against perturbations. Therefore, crystalline symmetry is needed to protect the 2D DPs.

The organization of this paper is as follows. After this introductory section, Sec.~\ref{theory} describes the general framework to classify stable 2D DSM with $\mathcal{PT}$ symmetry. Section~\ref{model}
presents a magnetic tight-binding model which supports symmetry-protected 2D DPs, and the symmetry-breaking phases. Section IV presents discussion on the possible material venues for 2D AFM DSMs, and concludes this paper. Some auxiliary materials are relegated
to Appendixes.

\section{General Theory}
\label{theory}

The basic mechanism for the avoidable crossing of energy bands is to let the bands have different symmetry representations~\cite{yang2014,young2015,fang2015}. We start with classifying stable DSMs of 2D SOC systems in the presence of $\mathcal{PT}$ and crystalline symmetry such as rotational and mirror symmetries. The generic Hamiltonian has the form $\mathcal{H}(\mathbf{k})=\sum_{i,j=0}^3d_{ij}(\mathbf{k})\sigma_i\otimes\tau_j$, where $\sigma_i$ and $\tau_i$ ($i=1,2,3$) are Pauli matrices acting on the spin and orbital, respectively. $\sigma_0$ and $\tau_0$ are $2\times2$ identity matrices. $d_{ij}(\mathbf{k})$ are real functions of $\mathbf{k}$. The invariance of the system will set many terms to be zero or nonindependent, which reduces to the effective Hamiltonian in Eq.~(\ref{minimal}).

\subsection{Case A. $\mathcal{T}$, $\mathcal{P}$ conserved}
The $\mathcal{T}$ operator is $\mathcal{T}=i\sigma_2\mathcal{K}$, where $\mathcal{K}$ is complex conjugation. $\mathcal{H}(-\mathbf{k})=\mathcal{T}\mathcal{H}(\mathbf{k})\mathcal{T}^{-1}$. $\mathcal{P}$ is independent of the spin-rotation, where its specific form is determined by $\mathcal{P}^\dag\mathcal{P}=1$, $[\mathcal{T},\mathcal{P}]=0$, and $(\mathcal{PT})^2=-1$. Thus $\mathcal{P}=\pm\tau_0$, $\pm\tau_3$, or $\pm\tau_1$. $\mathcal{H}(-\mathbf{k})=\mathcal{P}\mathcal{H}(\mathbf{k})\mathcal{P}^{-1}$.
If $\mathcal{P}=\pm\tau_1$, $\Gamma^{(1,2,3,4,5)}=(\sigma_3\otimes\tau_3,\tau_2,\sigma_1\otimes\tau_3,\sigma_2\otimes\tau_3,\tau_1)$,
and $d_a(\mathbf{k})=-d_a(-\mathbf{k})$ for $a=1,\ldots,4$, $d_5(\mathbf{k})=d_5(-\mathbf{k})$. All $d_a(\mathbf{k})=0$ satisfied simultaneously leads to DPs, however, the number of equations is larger than the number of variables $k_x$, $k_y$ and the external parameter $m$. Therefore, the band crossing will not happen at generic $\mathbf{k}$. The crystalline symmetry will further set constraints on $d_a(\mathbf{k})$. In 2D, the relevant symmetries are $C_{2\hat{x}}$, $C_{2\hat{y}}$, $M_{\hat{z}}$, and $C_{n\hat{z}}$.

\begin{table}[b]
\caption{The classification table of 2D DSMs for \emph{case A} with both $\mathcal{T}$ and $\mathcal{P}$ symmetries. $\ell=2,4,6$. The DPs are induced by the topological band crossing at the BZ boundary.}
\begin{center}\label{table1}
\renewcommand{\arraystretch}{1.5}
\begin{tabular*}{3.4in}
{@{\extracolsep{\fill}}cccc}
\hline
\hline
Symmetry & $\mathcal{P}$ & $\mathcal{T}$ &  Dispersion
\\
\hline
$C_{2\hat{x}}=i\sigma_3\otimes\tau_3$ & $\pm\tau_1$ & $i\sigma_2\mathcal{K}$ &  Linear
\\
$M_{\hat{z}}=\sigma_3\otimes\tau_2$ & $\pm\tau_1$ & $i\sigma_2\mathcal{K}$ &  Linear
\\
$C_{\ell\hat{z}}=e^{i\frac{\pi}{\ell}\sigma_3}\otimes\tau_3$ & $\pm\tau_1$ & $i\sigma_2\mathcal{K}$ &  Linear
\\
$C_{6\hat{z}}=e^{i\frac{\pi}{2}\sigma_3}\otimes\tau_3$ & $\pm\tau_1$ & $i\sigma_2\mathcal{K}$ &  Cubic
\\
\hline
\hline
\end{tabular*}
\end{center}
\end{table}

\emph{$C_{2\hat{x}}$ symmetry.} The invariance of the system under $\pi$ rotation along the $x$ axis requires $C_{2\hat{x}}\mathcal{H}(k_x,k_y)C_{2\hat{x}}^{-1}=\mathcal{H}(k_x,-k_y)$. $k_y=0$ along the $k_x$ axis, $[C_{2\hat{x}},\mathcal{H}(k_x,0)]=0$. Therefore we can choose a basis to make both $C_{2\hat{x}}$ and $\mathcal{H}(k_x)$ diagonal. In such a basis, the explicit forms of $C_{2\hat{x}}$ is obtained by the constraint $[\mathcal{T},C_{2\hat{x}}]=0$. Therefore $C_{2\hat{x}}=\text{diag}[\alpha_p,\alpha_q,\alpha_p^*,\alpha_q^*]$, where $\alpha_p=\exp[i\pi(p+\frac{1}{2})]$ with $p=0,1$. Thus we have $C_{2\hat{x}}=i\sigma_3$, or $i\sigma_3\otimes\tau_3$.
With the explicit representations of $C_{2\hat{x}}$ and $\mathcal{P}$ operators, we can get the symmetry allowed forms of $d_a(\mathbf{k})$. Take $\mathcal{P}=\pm\tau_1$ for example, when $C_{2\hat{x}}=i\sigma_3\otimes\tau_3$, to the leading order in $\mathbf{k}$, the symmetry constraints leads to
$d_{1,2,3,4,5}(\mathbf{k})=(v_1k_x,v_2k_y,v_3k_y,v_4k_y,v_5k_xk_y)$,
where $v_i$ are coefficients. This is nothing but the 2D DSM with anisotropic linear dispersions. Considering the full periodic structure of the BZ, the time-reversal-invariant momentum (TRIM) $(\pi,0)$, $(0,\pi)$ and $(\pi,\pi)$ are possible locations of the 2D DPs. The $(0,0)$ point is simply excluded because the symmetry-protected DPs are not compatible with threefold rotation, while the space groups admit the 4D representations at $(0,0)$ contain $C_{3\hat{z}}$~\cite{young2012,bradley1972}. Similarly, when $C_{2\hat{x}}=i\sigma_3$,
$d_{1,2,3,4,5}(\mathbf{k})=(v_1k_x,v_2k_x,v_3k_y,v_4k_y,m+u_1k_x^2+u_2k_y^2)$, where $v_i$, $u_i$ and $m$ are constants. This is not DSM Hamiltonian. However, it is worth to mention that critical DPs indeed exist at TRIM at the transition between QSH and NI~\cite{murakami2007} as shown in Fig.~\ref{fig4}(b).
At TRIM point, $\mathbf{k}$ and $-\mathbf{k}$ are equivalent, and all odd functions in $\mathcal{H}(\mathbf{k})$ vanishes. By tuning $m$, $d_5(\mathbf{k},m)=0$ is satisfied.

The classification of 2D DSMs for \emph{case A} is shown in Table~\ref{table1}. The details on 2D DPs protected by $M_{\hat{z}}$ and $C_{n\hat{z}}$ are discussed in Appendix~\ref{appendix_A}. The stable DPs exist at TRIM of the BZ boundary and are protected by the lattice symmetry, including symmorphic and nonsymmorphic symmetries~\cite{young2015}. In nonsymmorphic symmetric systems with two sublattices,
$\mathcal{P}=\pm\tau_1$ is the inversion operator with sublattices interchanged~\cite{yang2014}, which is
realized by the inversion with respect to a lattice site followed by a fraction translation.

\begin{figure}[t]
\begin{center}
\includegraphics[width=3.3in]{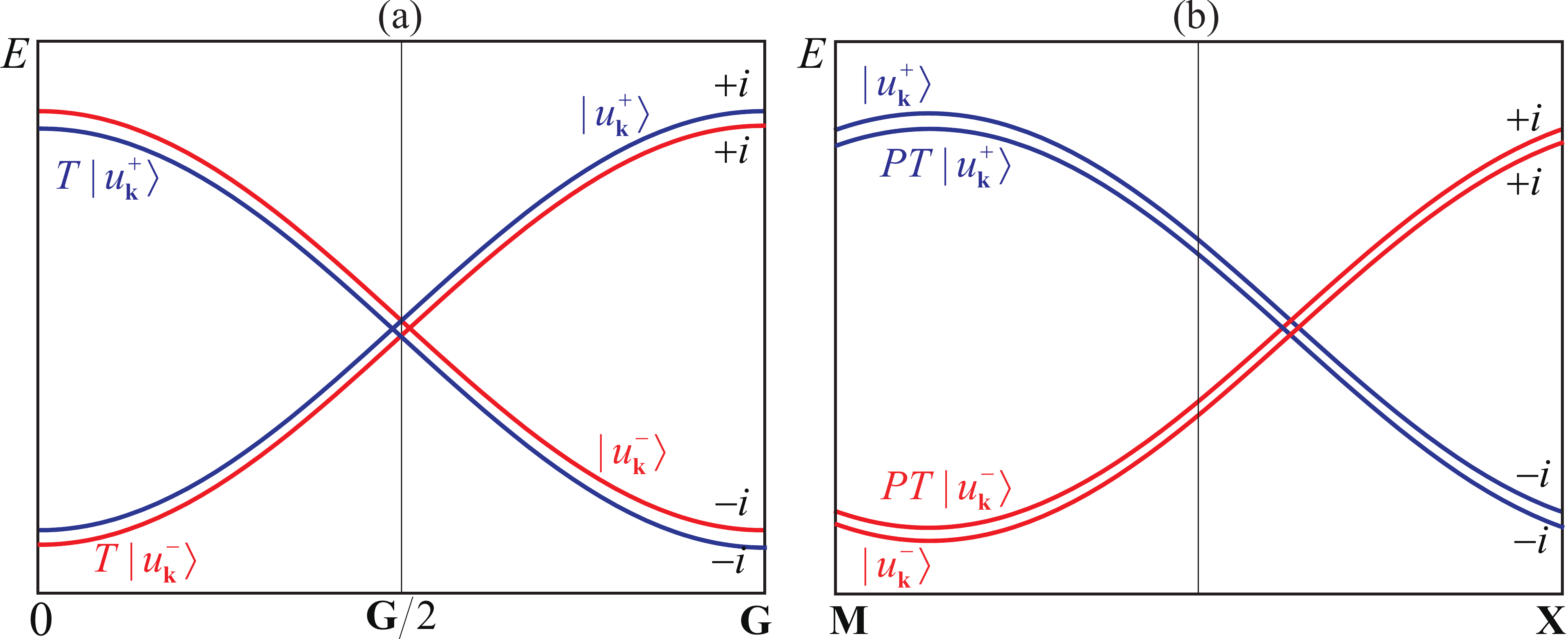}
\end{center}
\caption{(color online). A nonsymmorphic symmetry $\{g|\mathbf{t}\}$ ensures band crossing on a $g$ invariant line at the BZ boundary. The doubly-degenerate bands are artificially split for clarity. (a) With $\mathcal{P}$ and $\mathcal{T}$, and $\mathcal{P}=\tau_1$, Bloch states  $(|u^+_{\mathbf{k}}\rangle,\mathcal{T}|u^-_{\mathbf{k}}\rangle)$ and $(|u^-_{\mathbf{k}}\rangle,\mathcal{T}|u^+_{\mathbf{k}}\rangle)$ carry different representations of $\{g|\mathbf{t}\}$ ($C_{2\hat{x}}$ for example) and cross at TRIM. (b) Without $\mathcal{P}$ and $\mathcal{T}$, but with combined $\mathcal{PT}$, $(|u^+_{\mathbf{k}}\rangle,\mathcal{PT}|u^+_{\mathbf{k}}\rangle)$ and $(|u^-_{\mathbf{k}}\rangle,\mathcal{PT}|u^-_{\mathbf{k}}\rangle)$ have different eigenvalues of $\{C_{2\hat{x}}|\frac{1}{2}\frac{1}{2}\}$ and cross avoidably.}
\label{fig1}
\end{figure}

\subsection{Case B. $\mathcal{T}$, $\mathcal{P}$ broken}

The anti-unitary $\mathcal{PT}$ symmetry reverses spins and keeps $\mathbf{k}$ invariant. Without loss of generality, in this section we set $\mathcal{PT}=i\sigma_2\mathcal{K}$. With $[\mathcal{PT},\mathcal{H}(\mathbf{k})]=0$, one has $\mathcal{H}(\mathbf{k})=d_1\tau_1+d_2\sigma_3\otimes\tau_2+d_3\sigma_1\otimes\tau_2+d_4\sigma_2\otimes\tau_2+d_5\tau_3$.
The crystalline symmetries fall into two classes: symmorphic and nonsymmorphic space group symmetries, the operations of which is denoted as $\{g|\mathbf{t}\}$ and can be constructed by the point group operations $g$ with translation $\mathbf{t}$ that are a full and a fraction of a Bravais lattice vector, respectively. In 2D, the representative symmorphic group symmetries are $\{C_{2\hat{x}}|00\}$, $\{C_{2\hat{y}}|00\}$,  $\{M_{\hat{z}}|00\}$ and $C_{n\hat{z}}$. The typical 2D nonsymmorphic group operations include screw rotation $\{C_{2\hat{x}}|\mathbf{t}\}$, $\{C_{2\hat{y}}|\mathbf{t}\}$, glide mirror lines $\{M_{\hat{x}}|\mathbf{t}\}$, $\{M_{\hat{y}}|\mathbf{t}\}$, and glide mirror plane $\{M_{\hat{z}}|\mathbf{t}\}$, where $\mathbf{t}$ is a half-translation which satisfying $g\mathbf{t}=\mathbf{t}$ and $e^{i\mathbf{G}\cdot\mathbf{t}}=-1$ for odd reciprocal lattice vector $\mathbf{G}$. Explicitly, in units of Bravais lattice constant, $\mathbf{t}=(\frac{1}{2},0),(0,\frac{1}{2}),(\frac{1}{2},\frac{1}{2})$. In the following, we will show that stable 2D DPs can only exist in the presence of nonsymmorphic symmetries.

(i). For the \emph{symmorphic group}, we take $\{M_{\hat{z}}|00\}$ for example, $\mathbf{k}$ is invariant and $(\sigma_1,\sigma_2,\sigma_3)\rightarrow(-\sigma_1,-\sigma_2,\sigma_3)$. The symmetry constraints on $d_a(\mathbf{k})$ is $d_{3,4}(\mathbf{k})=0$. To the lowest order in $\mathbf{k}$, the general form is $d_{a}(\mathbf{k})=m_a+v_a^xk_x+v_a^yk_y$ for $a=1,2,5$, and $m_a,v_a^x,v_a^y$ are coefficients. The DPs are not guaranteed and can only be generated by fune tuning. Even with the extra $\{C_{2\hat{x}}|00\}$ or $\{C_{2\hat{y}}|00\}$ symmetry, it only constrains $d_2(\mathbf{k})=v_2^yk_y$, still the DPs are not guaranteed.

\begin{table}[b]
\caption{The classification table of 2D DSMs for \emph{case B} with $\mathcal{T}$ and $\mathcal{P}$ broken. The nonsymmorphic symmetries with corresponding half-translation $\mathbf{t}$ are indicated.}
\begin{center}\label{table2}
\renewcommand{\arraystretch}{1.5}
\begin{tabular*}{3.4in}
{@{\extracolsep{\fill}}ccc}
\hline
\hline
Symmetry & Half-translation $\mathbf{t}$ & Dispersion
\\
\hline
$\{C_{2\hat{x}}|\mathbf{t}\}$ & $\mathbf{t}=(0,\frac{1}{2})$, $(\frac{1}{2},\frac{1}{2})$ & Linear
\\
$\{C_{2\hat{y}}|\mathbf{t}\}$ & $\mathbf{t}=(\frac{1}{2},0)$, $(\frac{1}{2},\frac{1}{2})$ & Linear
\\
$\{M_{\hat{x}}|\mathbf{t}\}$ & $\mathbf{t}=(\frac{1}{2},0)$, $(\frac{1}{2},\frac{1}{2})$ & Linear
\\
$\{M_{\hat{y}}|\mathbf{t}\}$ & $\mathbf{t}=(0,\frac{1}{2})$, $(\frac{1}{2},\frac{1}{2})$ & Linear
\\
\hline
\hline
\end{tabular*}
\end{center}
\end{table}

(ii). For the \emph{nonsymmorphic group}, we take $\{C_{2\hat{x}}|\frac{1}{2}\frac{1}{2}\}$ for example.
Under $\{C_{2\hat{x}}|\frac{1}{2}\frac{1}{2}\}$, $(x,y)\rightarrow(x+\frac{1}{2},-y+\frac{1}{2})$, $(k_x,k_y)\rightarrow(k_x,-k_y)$ and $(\sigma_1,\sigma_2,\sigma_3)\rightarrow(\sigma_1,-\sigma_2,-\sigma_3)$.
$\{C_{2\hat{x}}|\frac{1}{2}\frac{1}{2}\}\mathcal{H}(k_x,k_y)\{C_{2\hat{x}}|\frac{1}{2}\frac{1}{2}\}^{-1}=\mathcal{H}(k_x,-k_y)$.
Thus $\{C_{2\hat{x}}|\frac{1}{2}\frac{1}{2}\}^2=-e^{-ik_x}$ is just a full-translation along $x$ axis with a $2\pi$ rotation of spins.
$\mathcal{H}(\mathbf{k})$ is invariant under $\{C_{2\hat{x}}|\frac{1}{2}\frac{1}{2}\}$ along $k_y=0,\pi$ lines. Therefore both $\{C_{2\hat{x}}|\frac{1}{2}\frac{1}{2}\}$ and $\mathcal{H}(k_x)$ can be chosen as diagonal.
The exact representation of $\{C_{2\hat{x}}|\frac{1}{2}\frac{1}{2}\}$ operator is obtained by noticing
\begin{equation}\label{anticommutation} \{C_{2\hat{x}}|\frac{1}{2}\frac{1}{2}\}\mathcal{PT}=e^{-ik_x}e^{-ik_y}\mathcal{PT}\{C_{2\hat{x}}|\frac{1}{2}\frac{1}{2}\}.
\end{equation}
Thus $\{C_{2\hat{x}}|\frac{1}{2}\frac{1}{2}\}=ie^{-ik_x/2}\sigma_3$, or $ie^{-ik_x/2}\sigma_3\otimes\tau_3$ when $k_y=0$; and $\{C_{2\hat{x}}|\frac{1}{2}\frac{1}{2}\}=ie^{-ik_x/2}\tau_3$ when $k_y=\pi$. However, there is only one case that stable band crossing and Dirac Hamiltonian can appear. Namely, $\{C_{2\hat{x}}|\frac{1}{2}\frac{1}{2}\}=ie^{-ik_x/2}\tau_3$ along $k_y=\pi$ line, only $d_5(k_x,\pi)$ term is present. If $d_5(k_x,\pi)=0$ at $\mathbf{k}_0=(k_{x0},\pi)$, around $\mathbf{k}_0$ point, to the leading order, the effective Hamiltonian is Dirac like and expanded as $d_{1,2,3,4,5}(\mathbf{k})=(v_1k_y,v_2k_y,v_3k_y,v_4k_y,v_5k_x)$, where $v_i$ are real. There is an alternative way to see why topological band crossing can only exist along $k_y=\pi$ line. The two Bloch states $|u^+_{\mathbf{k}}\rangle$, $|u^-_{\mathbf{k}}\rangle$ and their $\mathcal{PT}$ partners $\mathcal{PT}|u^+_{\mathbf{k}}\rangle$, $\mathcal{PT}|u^-_{\mathbf{k}}\rangle$ are eigenstates of $\{C_{2\hat{x}}|\frac{1}{2}\frac{1}{2}\}$, which is $\pm ie^{-ik_x/2}$. $|u^\pm_{\mathbf{k}}\rangle$ has the same energy as its $\mathcal{PT}$ partner. To have a stable band crossing, ($|u^+_{\mathbf{k}}\rangle,\mathcal{PT}|u^+_{\mathbf{k}}\rangle$) and ($|u^-_{\mathbf{k}}\rangle,\mathcal{PT}|u^-_{\mathbf{k}}\rangle$) should carry different representations of $\{C_{2\hat{x}}|\frac{1}{2}\frac{1}{2}\}$ as shown in Fig.~\ref{fig1}(b). This is only possible as $\{C_{2\hat{x}}|\frac{1}{2}\frac{1}{2}\}=ie^{-ik_x/2}\tau_3$ along $k_y=\pi$.

As shown in Table~\ref{table2}, the stable DPs can indeed exist in certain AFM semimetals respecting $\mathcal{PT}$ symmetry. The DPs in both $\mathcal{T}$-invariant and $\mathcal{T}$-broken semimetals are generated by the avoidable crossing of energy bands with distinct representations, which are protected by nonsymmorphic symmetries. However, different from the $\mathcal{T}$-invariant semimetals where the DPs reside at TRIMs of the BZ boundary, here the DPs in $\mathcal{T}$-broken systems can only appear at the BZ boundary but not at TRIMs in general. Moreover, the glide mirror plane symmetry $M_{\hat{z}}$ can not protect the DPs with $\mathcal{T}$-broken as listed in Appendix~\ref{appendix_B}, contrary to its protection of DPs with $\mathcal{T}$. More importantly, the DPs in the $\mathcal{T}$-broken AFM DSMs are locally permitted by crystalline symmetries, while the DPs in $\mathcal{T}$-invariant DSMs are essential and cannot be gapped without lowering the specific space group symmetries.

\section{Model}
\label{model}

\subsection{Tight-binding model}

\begin{figure}[t]
\begin{center}
\includegraphics[width=3.3in]{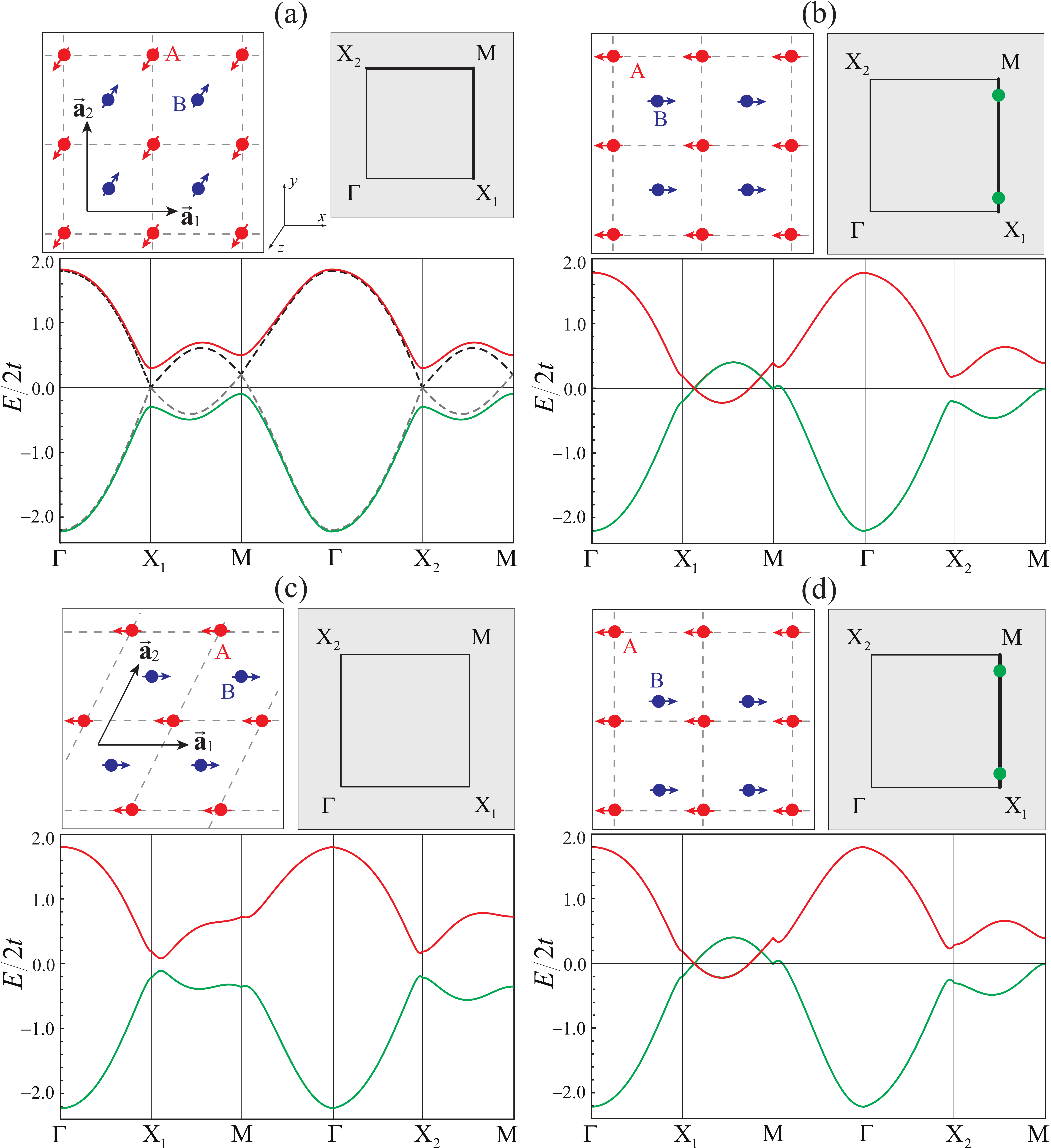}
\end{center}
\caption{(color online). Energy band with DPs protected by nonsymmorphic symmetries in an AFM lattice respecting $\mathcal{PT}$. The DPs are marked as green dots in the BZ. (a) DPs are gapped and cannot be protected by $\{C_{2\hat{x}}|\frac{1}{2}0\}$ and $\{C_{2\hat{y}}|0\frac{1}{2}\}$ when $\hat{\mathbf{n}}=\hat{z}$. (b) Two DPs located along $X_1$-$M$ line are protected by $\{M_{\hat{x}}|\frac{1}{2}0\}$ solely when $\hat{\mathbf{n}}=\hat{x}$. (c) Distortion in the $\langle11\rangle$ direction eliminates $\{M_{\hat{x}}|\frac{1}{2}0\}$ but keeps $\{M_{\hat{z}}|\frac{1}{2}\frac{1}{2}\}$, gapping the DPs. (d) Alternatively, $B$ sites shifted in $y$ direction breaks $\{M_{\hat{z}}|\frac{1}{2}\frac{1}{2}\}$, but DPs still remain protected.}
\label{fig2}
\end{figure}

Now we study a tight-binding model for a 2D AFM DSM to illustrate the nonsymmorphic symmetries protected DPs listed in Table~\ref{table2}. For direct comparison with $\mathcal{T}$ invariant case, we adopted the lattice similar to Ref.~\onlinecite{young2015} as shown in Fig.~\ref{fig2}(a), the system has a layered AFM structure with two atoms in one unit cell, denoted as $A$ and $B$. The lattice vectors are denoted as $\vec{\mathbf{a}}_1=(1,0,0)$, $\vec{\mathbf{a}}_2=(0,1,0)$. In each unit cell, the $A$ and $B$ atoms are shifted along the $z$ axis with the position $r_A=(-\frac{1}{4},-\frac{1}{4},-\frac{c}{2})$ and $r_B=(\frac{1}{4},\frac{1}{4},\frac{c}{2})$. Without AFM ordering, this structure is in space group No.~129 ($P4/nmm$). Each lattice site contains a $d_{z^2}$ orbital, where the square pyramidal crystal field splits $d_{z^2}$ from other $d$ orbitals. The Hamiltonian is
\begin{eqnarray}\label{H}
\mathcal{H} &=& t\sum\limits_{\langle
ij\rangle}c_{i}^{\dag}c_{j}+\sum\limits_{\langle\langle
ij\rangle\rangle}c_{i}^{\dag}\left[t_2+
i\lambda_{\text{so}}(\hat{\mathbf{d}}_1\times\hat{\mathbf{d}}_2)\cdot\mathbf{s}\right]c_{j}
\nonumber
\\
&&+\Delta\sum\limits_i\xi_ic_i^{\dag}\mathbf{s}\cdot\hat{\mathbf{n}}c_i,
\end{eqnarray}
where $\langle ij\rangle$ and $\langle\langle ij\rangle\rangle$
denote the nearest and next-nearest neighbor sites, respectively. $\lambda_{\text{so}}$ is SOC which involves spin-dependent next-nearest neighbor hopping, where $\hat{\mathbf{d}}_1$ and $\hat{\mathbf{d}}_2$ are unit vectors along two nearest neighbor bonds hopping from site $j$ to $i$~\cite{kane2005b}. $\mathbf{s}$ describes the electron spins. The third term is the staggered Zeeman term ($\xi_i=\pm1$) which describes the AFM ordering (along $\hat{\mathbf{n}}$ direction).

The symmetry of the system depends on the AFM order direction. In Fig.~\ref{fig2}(a), if $\hat{\mathbf{n}}=\hat{z}$, the system breaks $\mathcal{T}$ and $\mathcal{P}$ but respects $\mathcal{PT}$. However, from the energy bands, the screw axes $\{C_{2\hat{x}}|\frac{1}{2}0\}$ and $\{C_{2\hat{y}}|0\frac{1}{2}\}$ can not protect the DPs, consistent with the analysis in Table~\ref{table2}. This is quite different from the $\mathcal{T}$-, $\mathcal{P}$-invariant case (dashed line), where the DPs at $X_1$, $M$ and $X_2$ are protected by $\{C_{2\hat{x}}|\frac{1}{2}0\}$ and $\{C_{2\hat{y}}|0\frac{1}{2}\}$~\cite{young2015}. In Fig.~\ref{fig2}(b), if $\hat{\mathbf{n}}=\hat{x}$, two DPs are located along the $X_1$-$M$ line, and they are at different energies in the presence of $t_2$ term. The Hamiltonian
\begin{eqnarray}
\mathcal{H}(\mathbf{k}) &=& 4t\tau_1\cos\frac{k_x}{2}\cos\frac{k_y}{2}+2t_2(\cos k_x+\cos k_y)+
\nonumber
\\
&&2\lambda_{\text{so}}\sigma_2\otimes\tau_3\sin k_x+(\Delta-2\lambda_{\text{so}}\sin k_y)\sigma_1\otimes\tau_3.
\nonumber
\end{eqnarray}
The DPs are at $\mathbf{k}_1$=$(\pi,k_{y0})$ and $\mathbf{k}_2$=$(\pi,\pi-k_{y0})$, where $\sin k_{y0}=\Delta/2\lambda_{\text{so}}$. These two inequivalent DPs are protected by $\{M_{\hat{x}}|\frac{1}{2}0\}$. This can be seen by examining the effective model near these points. Near $\mathbf{k}=\mathbf{k}_1$,
\begin{eqnarray}
\mathcal{H}(\mathbf{k}_1+\mathbf{q})
&=& -2t\cos\frac{k_{y0}}{2}\tau_1q_x-2\lambda_{\text{so}}\sigma_2\otimes\tau_3q_x
\nonumber
\\
&&-2\lambda_{\text{so}}\cos k_{y0}\sigma_1\otimes\tau_3q_y.
\end{eqnarray}
At $\mathbf{k}_1$, the symmetry $\mathcal{PT}=i\sigma_2\mathcal{K}\otimes\tau_1$ allows the mass terms $\tau_2$ and $\sigma_3\otimes\tau_3$. This is forbidden by $\{M_{\hat{x}}|\frac{1}{2}0\}=i\sigma_1\otimes\tau_3$, but is allowed by $\{C_{2\hat{y}}|0\frac{1}{2}\}=\sigma_2\otimes\tau_2$. There is a subtle point, it seems $\{M_{\hat{z}}|\frac{1}{2}\frac{1}{2}\}=\sigma_3\otimes\tau_1$ could also forbid the mass terms.
To further clarify the role of $\{M_{\hat{x}}|\frac{1}{2}0\}$ and $\{M_{\hat{z}}|\frac{1}{2}\frac{1}{2}\}$ in protecting the DPs, we distorted the lattice which eliminates $\{M_{\hat{x}}|\frac{1}{2}0\}$ and keeps $\{M_{\hat{z}}|\frac{1}{2}\frac{1}{2}\}$ in Fig.~\ref{fig2}(c). Such distortion adds a term to nearest-neighbor hopping $\mathcal{H}_1=t_3\sin(k_x/2)\sin(k_y/2)\tau_1$, which will gap the DPs at $\mathbf{k}_{1,2}$, consistent with that DPs are not protected by $\{M_{\hat{z}}|\frac{1}{2}\frac{1}{2}\}$. In Fig.~\ref{fig2}(d), $B$ site is shifted along $y$ axis, which breaks $\{M_{\hat{z}}|\frac{1}{2}\frac{1}{2}\}$ but keeps $\{M_{\hat{x}}|\frac{1}{2}0\}$. Such reduced symmetry allows a term $\mathcal{H}_2=\cos(k_x/2)\sin(k_y/2)\left(t_4\tau_1+t_5\tau_2\right)$, where the DPs located along the $X_1$-$M$ line remain protected. However, it is noted that these two inequivalent DPs always appear or disappear together.

\subsection{Symmetry breaking}

\begin{figure}[b]
\begin{center}
\includegraphics[width=3.3in]{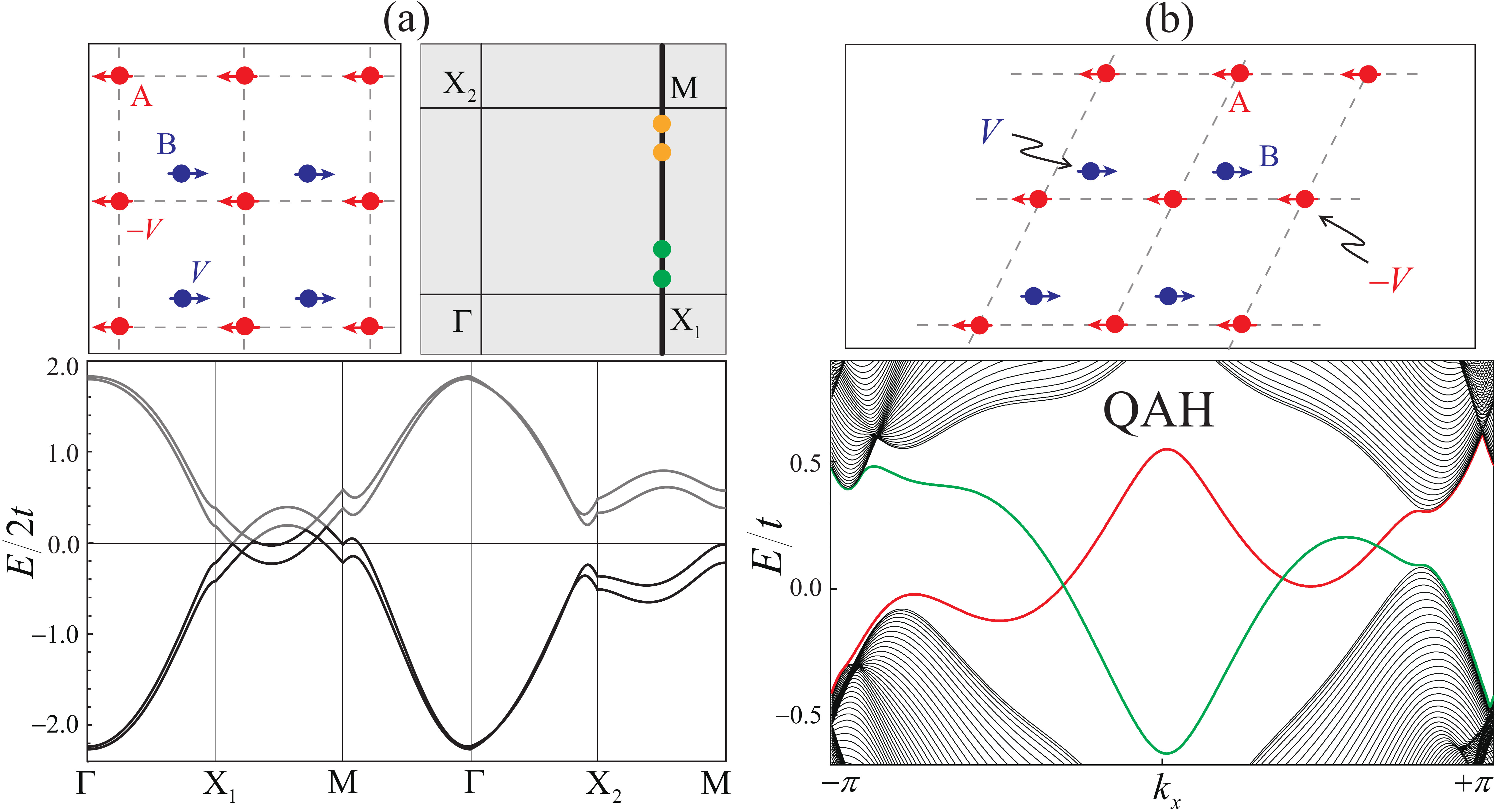}
\end{center}
\caption{(color online). Symmetry breaking phases. (a) Breaking $\mathcal{PT}$ by SIA, while preserving $\{M_{\hat{x}}|\frac{1}{2}0\}$, leads to Weyl points (marked as yellow and green dots) on the $k_x=\pi$ line. (b) Breaking both $\mathcal{PT}$ and $\{M_{\hat{x}}|\frac{1}{2}0\}$ results in the QAH insulator. In the edge spectrum, the green and red lines are edge states located at upper and lower edges, respectively.}
\label{fig3}
\end{figure}

Lowering the symmetry by external perturbations provides a toolbox to explore a wealth of topological phases. We start with the system as shown in Fig.~\ref{fig2}(d). The $\{M_{\hat{x}}|\frac{1}{2}0\}$ breaking by distortion in Fig.~\ref{fig2}(c) leads to a NI with gapped DPs. Displacing the $B$ sites along $[11]$ or $[1\bar{1}]$ always lead to NI phases. Quite different from $\mathcal{T}$-, $\mathcal{P}$-invariant 2D DSM where it lies at the boundary between QSH and NI~\cite{young2015}, the 2D AFM DSM lies in the NI phases deeply. This is simply because the edge of a physical system always breaks $\mathcal{PT}$ and leads to gapped edge states, even though a nontrivial bulk $Z_2$ index may be defined. Similarly, the structure inversion asymmetry (SIA) by applying an electric field along $z$ axis naturally breaks $\mathcal{PT}$ and adds a term $V\tau_3$, results in nondegenerate bands where DPs split into Weyl points as shown in Fig.~\ref{fig3}(a). Such Weyl points are located along the $k_x=\pi$ line and are not protected by symmetry. More interestingly, if both $\mathcal{PT}$ and $\{M_{\hat{x}}|\frac{1}{2}0\}$ are broken, a quantum anomalous Hall (QAH) state can be realized~\cite{dong2016} and electrically controllable as shown in Fig.~\ref{fig3}(b). Furthermore, with the proximity effect to an $s$-wave superconductor, the realization of a chiral topological superconductor with external tunability is expected~\cite{qi2010b,wang2015c,he2016,wang2016b}.

\section{Discussion and Conclusion}
\label{discussion}

The AFM long range order could exist in 2D. It should be interesting to study how the magnetic fluctuation and Coulomb interaction affect the stability of DPs, which is left to future work. In terms of realistic materials,
the actual existence of such phases in known materials remains an open question. However, similar to TI materials, the 2D magnetic DSMs may exist in materials with both strong SOC and specific space group symmetries. Furthermore, we comment on the possible candidate in the \emph{A}MnBi$_2$ family of compounds (\emph{A} = Ca, Sr, Eu, Yb)~\cite{kim2011,lee2013,guo2014,borisenko2015,petrovic2016,masuda2016}. The transport experiments have already shown the 2D Dirac fermion behaviors in the bulk materials~\cite{masuda2016,petrovic2016}. Take SrMnBi$_2$ for example, its bulk structure is in space group No.~139 ($I4/mmm$). For a monolayer of this material, the AFM order on Mn atoms is along the [001] direction which breaks both $\mathcal{T}$ and $\mathcal{P}$ whereas $\mathcal{PT}$ still holds. The Bi atoms form two layers of square lattice which is similar to the structure in Fig.~\ref{fig2}(a), which determine most of the electronic structure~\cite{kim2011,lee2013,guo2014}. The nonsymmorphic symmetries are $\{C_{2\hat{x}}|\frac{1}{2}\frac{1}{2}\}$ and $\{C_{2\hat{y}}|\frac{1}{2}\frac{1}{2}\}$. Therefore, the 2D DPs if exist in this material, can be protected by these symmetries in principle. Unfortunately, this system shows \emph{massive} Dirac fermions around the Fermi level, with Dirac mass gap along the line $\Gamma$-$M$ and $M$-$X$~\cite{kim2011,lee2013}. In fact, it is similar to the model in Eq.~(\ref{H}) with AFM ordering $\hat{\mathbf{n}}$ along $[001]$ direction.

In summary, we show that the 2D Dirac fermions could exist in AFM semimetals with $\mathcal{PT}$ symmetry, where the nonsymmorphic space group symmetries play an essential role. The realistic AFM DSM materials remain unknown. However, considering the ongoing rapid progress in the field of 2D materials~\cite{novoselov2016}, together with the nonsymmorphic symmtries listed in Table~\ref{table1} as a guidance, we are thus optimistic about the material search of the 2D AFM DSMs.

{\it Note added}: Recently, we became aware of an independent work on a similar problem~\cite{young2016}. However, their approach to 2D magnetic DSMs is different from our results.

\begin{acknowledgments}
The author is grateful to Haijun Zhang, Biao Lian and Shoucheng Zhang for insightful discussions. This work is supported by the National Thousand-Young-Talents Program; the National Key Research Program of China under Grant No.~2016YFA0300703; the Natural Science Foundation of Shanghai under Grant No.~17ZR1442500; the Open Research Fund Program of the State Key Laboratory of Low-Dimensional Quantum Physics, through Contract No.~KF201606; and by Fudan University Initiative Scientific Research Program.
\end{acknowledgments}

\begin{appendix}

\section{$C_{n\hat{z}}$ and $M_{\hat{z}}$ symmetries in \emph{case A}}
\label{appendix_A}

The representations of the $\Gamma$ matrices depends on the crystalline symmetry. If $\mathcal{P}=\pm\tau_0$,
we get $\Gamma^{(1,2,3,4,5)}=(\tau_1,\sigma_3\otimes\tau_2,\sigma_1\otimes\tau_2,\sigma_2\otimes\tau_2,\tau_3)$
and $d_a(\mathbf{k})=d_a(-\mathbf{k})$ for $a=1,\ldots,5$. Similarly, if $\mathcal{P}=\pm\tau_3$,
$\Gamma^{(1,2,3,4,5)}=(\sigma_3\otimes\tau_1,\tau_2,\sigma_1\otimes\tau_1,\sigma_2\otimes\tau_1,\tau_3)$,
and $d_a(\mathbf{k})=-d_a(-\mathbf{k})$ for $a=1,\ldots,4$, $d_5(\mathbf{k})=d_5(-\mathbf{k})$.

\begin{figure}[b]
\begin{center}
\includegraphics[width=3.3in]{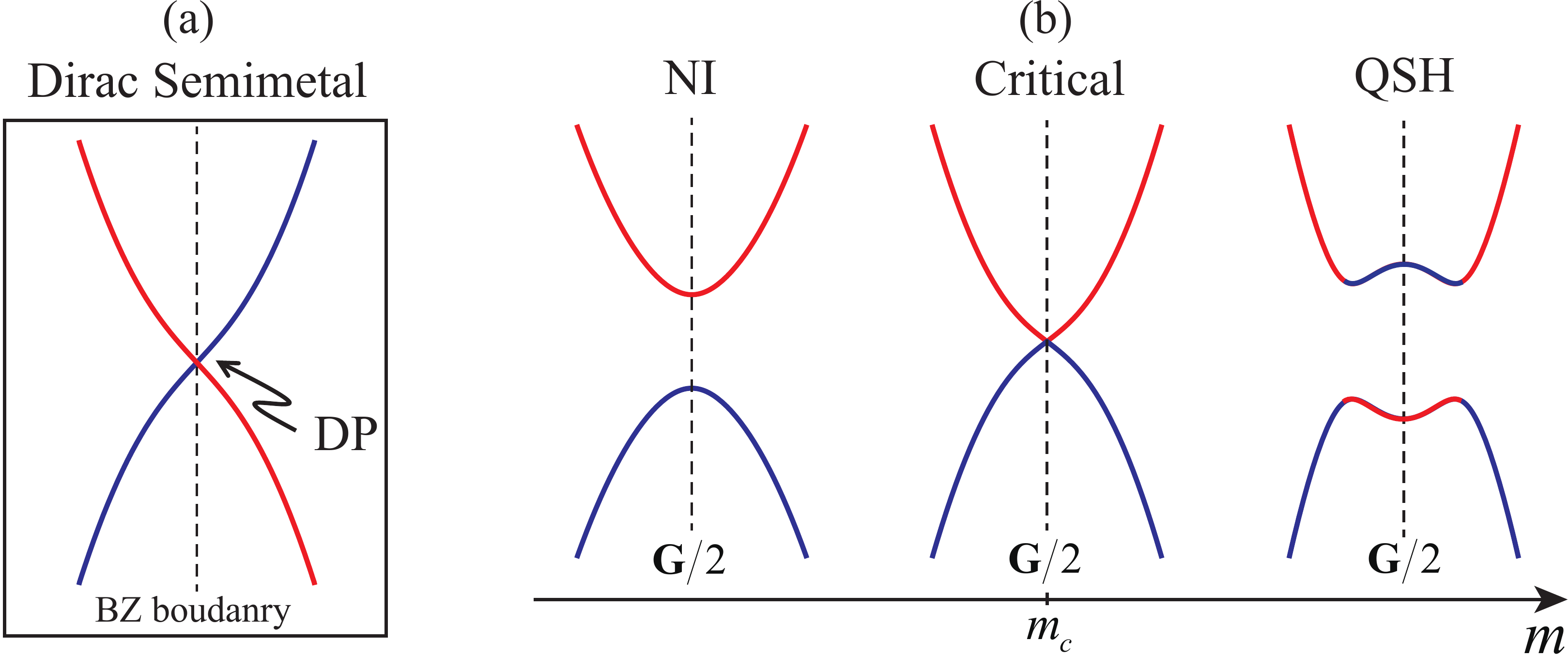}
\end{center}
\caption{(color online). (a) Nonsymmorphic symmetries protected 2D DSMs with DPs located at BZ boundary. (b) 2D DPs with accidental degeneracy appear at the phase transition between QSH and NI controlled by parameter $m$.}
\label{fig4}
\end{figure}

Here we give a short summary for the representations of $M_{\hat{z}}$ and $C_{n\hat{z}}$, and then give a detailed analysis for $C_{n\hat{z}}$ symmetry.

(i). \emph{$M_{\hat{z}}$ symmetry.} $M_{\hat{z}}$ is the mirror symmetry with respect to the $xy$ plane and requires $[M_{\hat{z}},\mathcal{H}(\mathbf{k})]=0$. It can be viewed as $M_{\hat{z}}=C_{2\hat{z}}\mathcal{P}$, where $C_{2\hat{z}}$ is the $\pi$ rotation along the $z$ axis. $M_{\hat{z}}$ follows the constraints $[\mathcal{T},M_{\hat{z}}]=0$, $M_{\hat{z}}^\dag M_{\hat{z}}=1$ and $M^2_{\hat{z}}=e^{i\phi}$.
The explicit form of $M_{\hat{z}}$ is (i) $M_{\hat{z}}=i\sigma_3\otimes\tau_1$, or $\sigma_3\otimes\tau_2$ when $|\mathcal{P}|=\tau_1$; (ii) $M_{\hat{z}}=i\sigma_3$, or $i\sigma_3\otimes\tau_3$ when $|\mathcal{P}|=\tau_0,\tau_3$. The stable DPs can appear with $M_{\hat{z}}=\sigma_3\otimes\tau_2$ and $\mathcal{P}=\pm\tau_1$.

(ii). \emph{$C_{n\hat{z}}$ symmetry.} $C_{n\hat{z}}$ is $n$-fold rotation along the $z$ axis, with $n=2,3,4,6$. The stable DPs can appear with $C_{\ell\hat{z}}$ ($\ell=2,4,6$) and $\mathcal{P}=\pm\tau_1$.

$C_{n\hat{z}}$ is the $n$-fold rotation along the $z$ axis, with $n=2,3,4,6$. In order to obtain the explicit form of $C_{n\hat{z}}$, we can choose a basis where $C_{n\hat{z}}$ is diagonal as $C_{n\hat{z}}=\text{diag}[u_{A}^{\uparrow},u_{B}^{\uparrow},u_{A}^{\downarrow},u_{B}^{\downarrow}]=\text{diag}[\alpha_p,\alpha_q,\alpha_p^*,\alpha_q^*]$, where $\alpha_p=\exp[i\frac{2\pi}{n}(p+\frac{1}{2})]$ with $p=0,1,\ldots,n-1$. The invariance of the system under $C_{n\hat{z}}$ requires
\begin{equation}\label{constraint}
C_{n\hat{z}}\mathcal{H}(k_+,k_-)C_{n\hat{z}}^{-1}=\mathcal{H}(k_+e^{i\frac{2\pi}{n}},k_-e^{-i\frac{2\pi}{n}}),
\end{equation}
where $k_\pm=k_x\pm ik_y$. Therefore, at the TRIM, the Hamiltonian commutes with $C_{n\hat{z}}$, i.e.,
\begin{equation}
[C_{n\hat{z}},\mathcal{H}(\mathbf{k}_{\text{TRIM}})]=0.
\end{equation}
This is different from $C_{2\hat{x}}$ and $M_{\hat{z}}$ cases, where invariant lines or plane exist in the BZ. We find that stable DPs can appear and are protected by $C_{2\hat{z}}$, $C_{4\hat{z}}$, $C_{6\hat{z}}$ together with $\mathcal{P}=\pm\tau_1$. The classification of 2D DSM protected by $C_{n\hat{z}}$ symmetry in \emph{case A} is listed in Table.~\ref{table1}. From Table~\ref{table1}, we can see that the $C_{4\hat{z}}$ case is consistent with the results obtained in Ref.~\onlinecite{young2015}. The $C_{2\hat{z}}$ case is not independent from $M_{\hat{z}}$ case in the main text, for in the presence of $\mathcal{P}$, $C_{2\hat{z}}=M_{\hat{z}}\mathcal{P}$.

Below we show the basic steps to see when the stable DPs can appear. As we discussed in the main text, the basic mechanism for the avoidable crossing of energy bands is to let the bands have different symmetry representations. The four Bloch states can be chosen as $|u^\uparrow_A\rangle$, $|u^\uparrow_B\rangle$, $|u^\downarrow_A\rangle$, and $|u^\downarrow_B\rangle$. Here $|u^\uparrow_A\rangle$ and $|u^\uparrow_B\rangle$ have the $C_{n\hat{z}}$ eigenvalues $u_{A}^{\uparrow}$ and $u_{B}^{\uparrow}$, respectively. There are two distinct cases,

(i). When $\mathcal{P}=\pm\tau_0,\pm\tau_3$, the inversion operation will not flip the orbitals, thus the energy $E_{A,\uparrow}=E_{A,\downarrow}$, and $E_{B,\uparrow}=E_{B,\downarrow}$ at $\mathbf{k}_{\text{TRIM}}$. Therefore, the four Bloch states are grouped into $(u_{A}^{\uparrow},u_{A}^{\downarrow})$ and $(u_{B}^{\uparrow},u_{B}^{\downarrow})$. In order to have a stable band crossing, one \emph{necessary} condition is that these two groups must have different eigenvalues of $C_{n\hat{z}}$. Namely, $(u_{A}^{\uparrow},u_{A}^{\downarrow})$ are different from $(u_{B}^{\uparrow},u_{B}^{\downarrow})$. Due to $[C_{n\hat{z}},\mathcal{H}({\mathbf{k}_{\text{TRIM}}})]=0$, both $C_{n\hat{z}}$ and $\mathcal{H}({\mathbf{k}_{\text{TRIM}}})$ can be diagonal. The only possible form is $\mathcal{H}({\mathbf{k}_{\text{TRIM}}})=d_5\tau_3$. However, $d_5$ is even under $\mathcal{P}$, thus $d_5=m$, where $m$ is the external parameters. In general, $m\neq0$, and Eq.~(\ref{constraint}) can not constrain $m$ to be zero, therefore, the stable Dirac points are not possible in this case.

(ii). Similarly, when $\mathcal{P}=\pm\tau_1$, inversion operation will switch the orbitals, therefore $E_{A,\uparrow}=E_{B,\downarrow}$, and $E_{B,\uparrow}=E_{A,\downarrow}$. Thus, $(u_{A}^{\uparrow},u_{B}^{\downarrow})$ must be different from $(u_{B}^{\uparrow},u_{A}^{\downarrow})$. This necessary condition rules out $C_{3\hat{z}}$. Moreover, we consider the only possible form of $\mathcal{H}$ at $\mathbf{k}_{\text{TRIM}}$ is $\mathcal{H}({\mathbf{k}_{\text{TRIM}}})=d_1\sigma_3\otimes\tau_3$. However, $d_1(\mathbf{k})$ is an odd function, thus $d_1$ vanishes. Furthermore, using Eq.~(\ref{constraint}) one can get the possible forms of $d_a(\mathbf{k})$. Here we take $C_{4\hat{z}}$ for example, from the above symmetry analysis, the only possible form is $C_{4\hat{z}}=e^{\pm i\frac{\pi}{4}\sigma_3}\otimes\tau_3$. At the TRIM, all odd functions $d_{1,2,3,4}$ vanish, while $d_5(\mathbf{k}_{\text{TRIM}})$ is an even function. But $C_{4\hat{z}}$ also force $d_5(\mathbf{k}_{\text{TRIM}})$ to be odd. Therefore $d_5$ vanishes at the TRIM, and DPs can appear the such TRIMs. At the TRIM, the Hamiltonian is expanded as $d_{1,2,3,4,5}(\mathbf{k})=(0,0,v k_x,v k_y,0)$. More interestingly, $C_{6\hat{z}}$ symmetry could support the cubic Dirac fermions.

Here we point out, in \emph{case A} with $\mathcal{P}$ symmetry, one can always construct the mirror symmetry $M_{\hat{n}_{\perp}}$ ($\hat{n}_\perp\perp\hat{z}$) from $C_{2\hat{n}_{\perp}}$ and $\mathcal{P}$ as $M_{\hat{n}_{\perp}}=C_{2\hat{n}_{\perp}}\mathcal{P}$. Here $\hat{n}_{\perp}$ can chosen to be $\hat{x}$ or $\hat{y}$. Therefore, $M_{\hat{n}_{\perp}}$ is not independent from $C_{2\hat{n}_{\perp}}$, and is not listed in Table~\ref{table1}. However, in \emph{case B} without $\mathcal{P}$, $M_{\hat{n}_{\perp}}$ is different from $C_{2\hat{n}_{\perp}}$, as we can see from Table~\ref{table1}.

\section{Nonsymmorphic symmetry with no DPs protection in \emph{case B}}
\label{appendix_B}

Here we list in Table~\ref{table3} the nonsymmorphic symmetries which cannot protect the DPs in \emph{case B}.

\begin{table}[htbp]
\caption{The nonsymmorphic symmetries which cannot protect the DPs in \emph{case B} with $\mathcal{T}$ and $\mathcal{P}$ broken.}
\begin{center}\label{table3}
\renewcommand{\arraystretch}{1.5}
\begin{tabular*}{3.4in}
{@{\extracolsep{\fill}}cccc}
\hline
\hline
Symmetry & half-translation $\mathbf{t}$ & DP protection
\\
\hline
$\{C_{2\hat{x}}|\mathbf{t}\}$ & $\mathbf{t}=(\frac{1}{2},0)$ & No
\\
$\{C_{2\hat{y}}|\mathbf{t}\}$ & $\mathbf{t}=(0,\frac{1}{2})$ & No
\\
$\{M_{\hat{x}}|\mathbf{t}\}$ & $\mathbf{t}=(0,\frac{1}{2})$ & No
\\
$\{M_{\hat{y}}|\mathbf{t}\}$ & $\mathbf{t}=(\frac{1}{2},0)$ & No
\\
$\{M_{\hat{z}}|\mathbf{t}\}$ & $\mathbf{t}=(\frac{1}{2},0), (0,\frac{1}{2}), (\frac{1}{2},\frac{1}{2})$ & No
\\
\hline
\hline
\end{tabular*}
\end{center}
\end{table}

\section{Electrically controllable QAH state}
\label{appendix_C}

\begin{figure}[b]
\begin{center}
\includegraphics[width=2.8in]{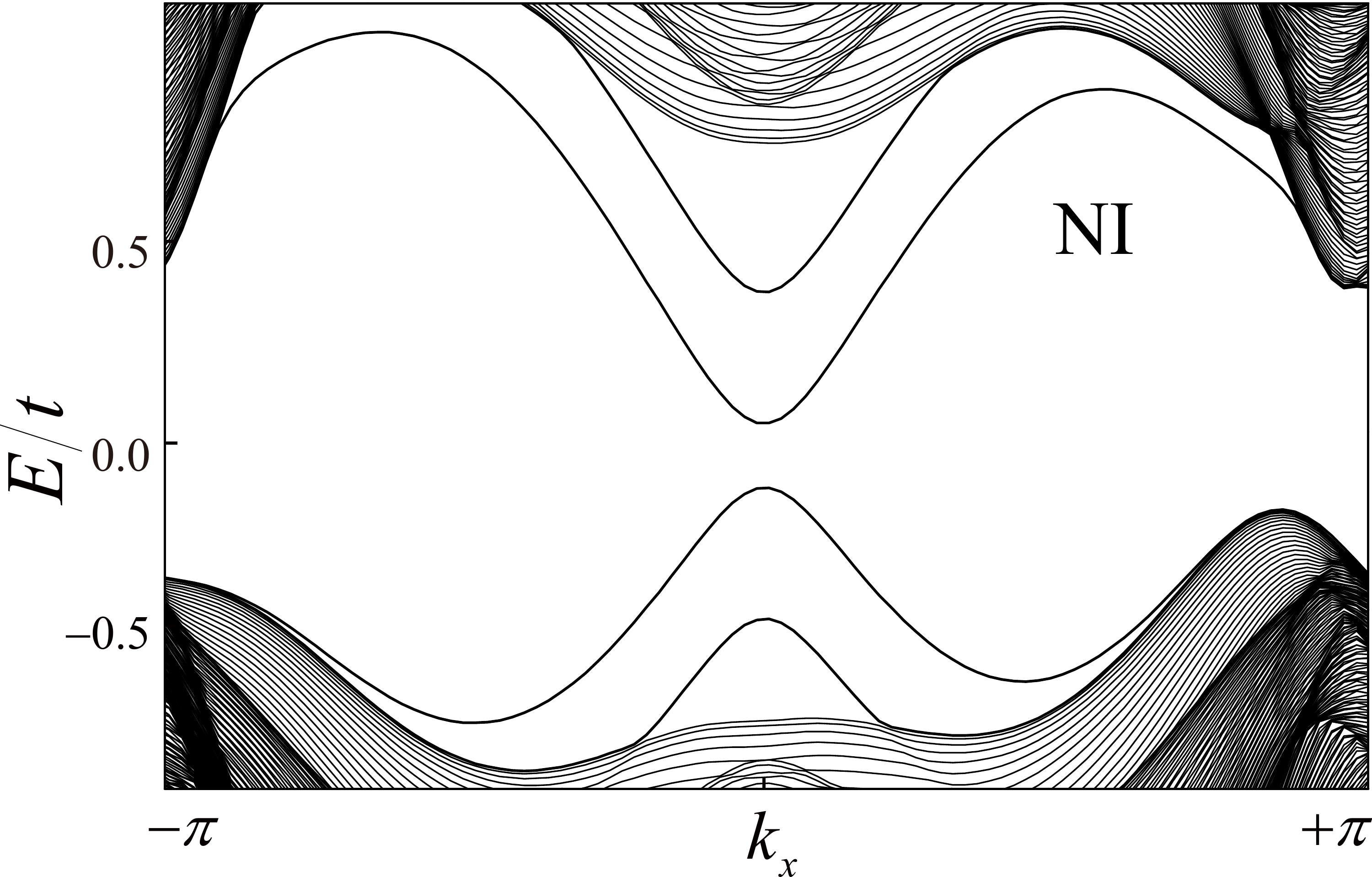}
\end{center}
\caption{The electrically tunable QAH state in both $\mathcal{PT}$ and $\{M_{\hat{x}}|\frac{1}{2}0\}$ broken AFM DSMs. The parameters $V=0.2$ and others are the same as in Fig.~\ref{fig3}(b) of main text, which leads to a NI phase.}
\label{fig5}
\end{figure}

The parameters in Fig.~\ref{fig3}(b) of main text is $t_2=-0.1$, $\lambda_{\text{so}}=0.5$, $\Delta=0.3$, $V=0.9$, and the four nearest-neighbor hopping are $t_{AB}=(1.5, 1.3, 1.3, 0.7)$. Meanwhile, the QAH state is electrically tunable, i.e., small $V$ leads to NI phase as shown in Fig.~\ref{fig5}.

\end{appendix}

\end{document}